\documentstyle[preprint,aps]{revtex}

\begin{document}

\input{epsf}
\draft
\preprint{{\bf MITH 98/9}\\ \\ }
\title{ 
COHERENT QED, GIANT RESONANCES AND $(e^{+}e^{-})$ PAIRS IN HIGH ENERGY 
NUCLEUS-NUCLEUS COLLISIONS \footnote{A preliminary, shortened version 
of this work appears in the Proceedings of the Conference CRIS$'98$, 
Catania, June $1998$.}
}
\author{{\bf R. Alzetta}, {\bf T.Bubba}, {\bf R. Le Pera}, {\bf G. 
Liberti}\footnote[5]{INFM, Unit\`a di Cosenza, Italia.},
{\bf G.Mileto}, {\bf D.Tarantino}}
\address{
Dipartimento di Fisica, Universit\'a della Calabria\\  
INFN, Cosenza, Italia. e-mail: Alzetta@fis.unical.it
}
\author{{\bf G. Preparata}}
\address{
Dipartimento di Fisica, Universit\`a di Milano\\ 
INFN, Milano, Italia. e-mail: Preparata@mi.infn.it
}
\maketitle
\begin{abstract}
We show that the coherent oscillations of the e.m. field induced by 
the collective quantum fluctuations of the nuclear matter field 
associated with the giant resonances, with frequencies 
$\omega_{A}\simeq 78A^{-\frac{1}{3}}$ MeV, give rise to a significant $(e^{+}e^{-})$
pair production in high energy Heavy Ion collisions. The approximate 
parameterless calculation of such yield is in good agreement with 
recent experimental observations.  
\end{abstract}

\vfill\eject

\section{Introduction}
Motivated by the search of a new state of hadronic matter, the 
Quark-Gluon-Plasma (QGP) predicted by some lattice Montecarlo 
studies \cite{Satz}, an impressive research program has been recently launched in 
the field of Heavy-Ion collisions at energies of a few hundred GeV 
per nucleon; for a recent review consult Ref.\cite{Drees}.
\par
\begin{figure}
 	\centering
     \mbox{\epsffile[0 0 400 430]{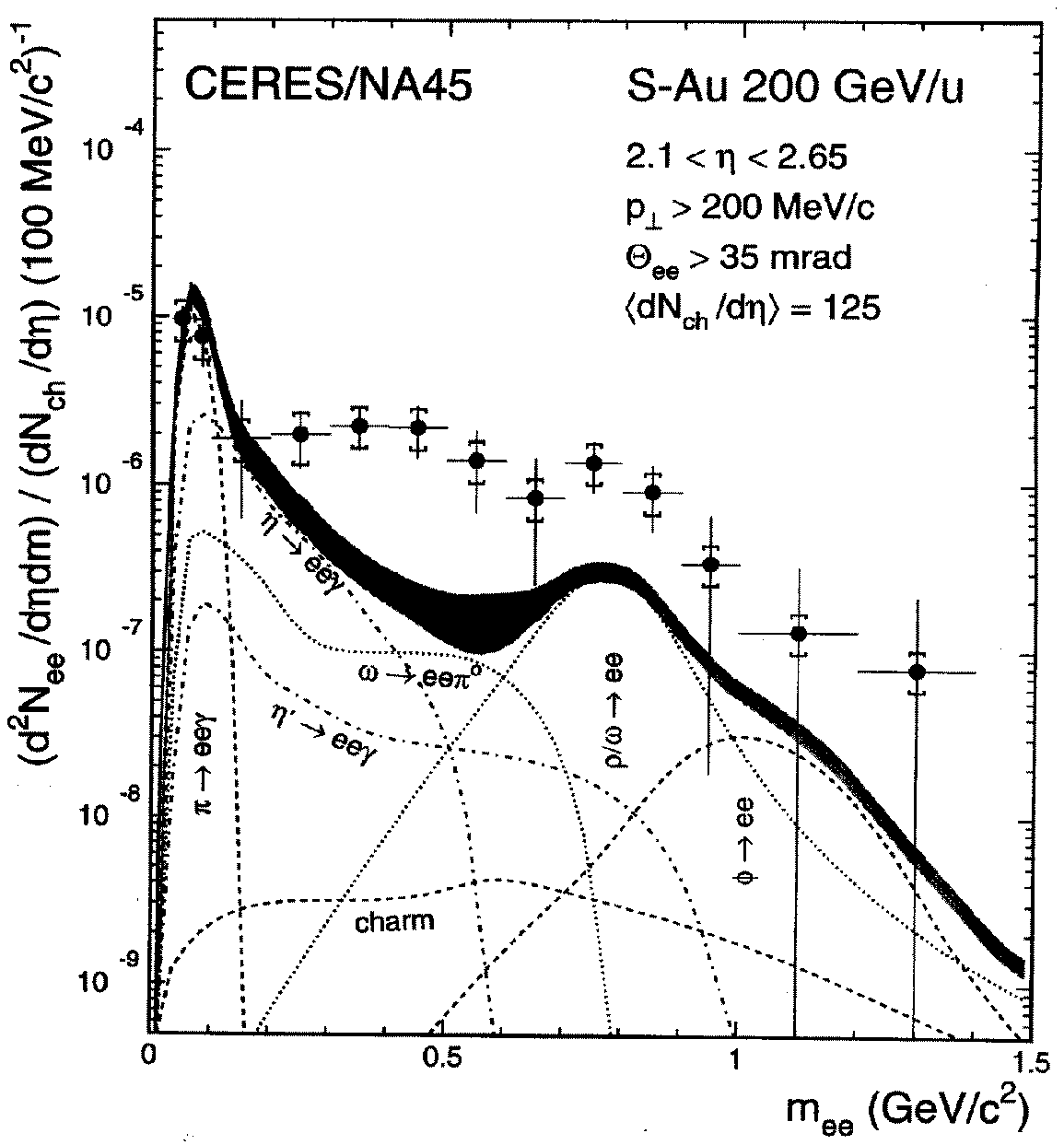}}
	 \vskip 1.5 cm
\caption{\em  Inclusive $e^{+}e^{-}$  mass spectra in 200 $GeV$/u S-Au 
collisions (CERES, Ref.3).}
\label{fig::S_AU200GEV}
\end{figure}

\begin{figure}
 	\centering
     \mbox{\epsffile[0 0 400 430]{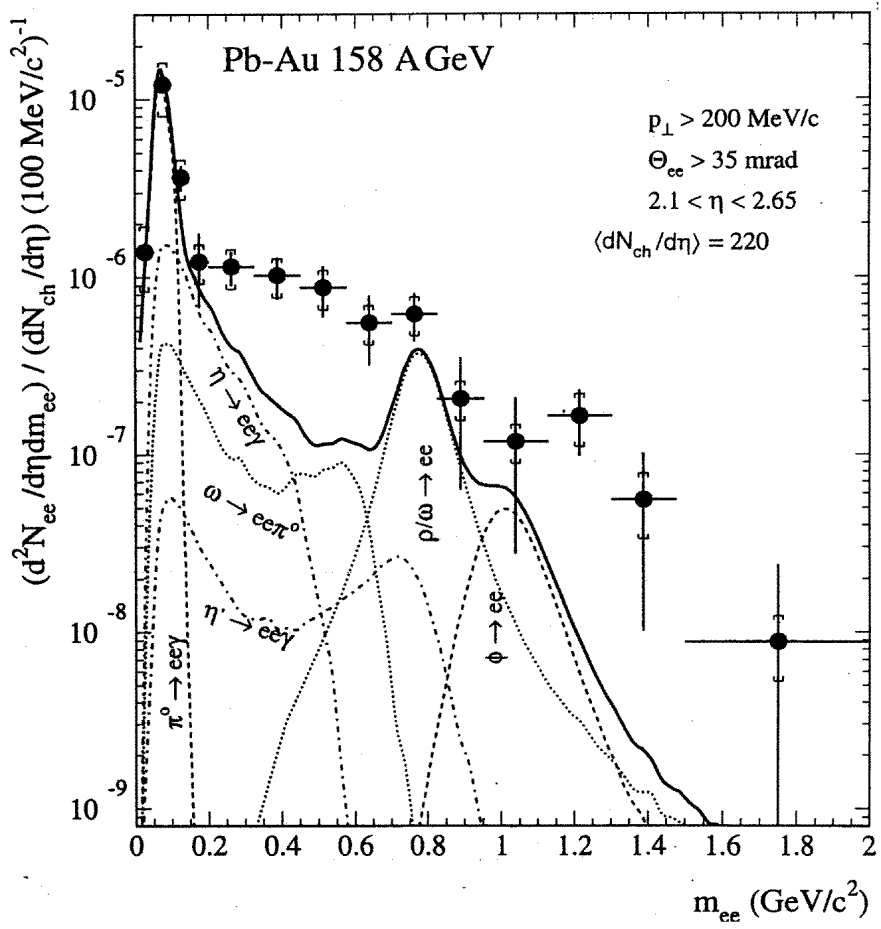}}
	 \vskip 1.5 cm
\caption{\em  Inclusive $e^{+}e^{-}$ mass spectra in 158 $GeV$/u Pb-Au 
collisions (CERES, Ref.5).}
\label{fig::PB_AU_RINORMA}
\end{figure}

 In particular three experiments have been conducted at the CERN-SPS 
in the last few years \cite{Aga1,Ullrich,Aga2} to measure the yield of low mass ($M< 2$ GeV)  
$(e^{+}e^{-})$ pairs in such collisions in search of some  
signal of the QGP. In Fig.\ref {fig::S_AU200GEV} and in Fig.\ref{fig::PB_AU_RINORMA} we 
report two typical observations of such experiments, showing a large 
yield of $(e^{+}e^{-})$ pairs in the mass region $(0.2<M<1.3$ GeV 
and $0.2<M<1.8$ GeV, respectively), well in excess of that predicted 
as arising from conventional sources. Furthermore in experiments with 
a high energy proton beam impinging on a heavy nucleus \cite{Drees} 
no such excess has been observed, suggesting that the source of the 
excess pairs lies in some peculiar aspect of the nucleus-nucleus 
interaction.
\par A wide consensus has formed so far around the hypotesis that the $(e^{+}e^{-})$
excess is due to some special manifestation of the QGP, and the 
important question is thus which peculiar QGP feature allows this 
hypotetical new state of matter to reveal its footprints in the form 
of abundant low mass $(e^{+}e^{-})$ pairs. As shown in Ref. 
\cite{Drees}, the situation is quite subtle and complicated, and the 
theory that appears now generally accepted \cite{liko} requires not 
only the production of an excess of vector mesons ($\rho, \omega, \phi$) 
over and above the normal yield in high energy pp-collisions, but also 
a substantial lowering of their masses, resulting from their 
propagation in a hot dense medium such as the QGP. And it is claimed 
\cite{liko}, that simulations with relativistic transport models 
reproduce the experimental findings.
\par While we have no technical objection to the deductions and simulations 
of Li,Ko and Brown \cite{liko}, we believe that the very notion of QGP, 
upon which this approach is based, is very arduous and difficult, and one 
that is quite unlikely to occur, due to the peculiar features of 
particle formation in high energy nucleon-nucleon scattering. A 
space-time analysis, in fact, of the process of particle production 
in high energy inelastic events \cite{Mueller} has made it rather 
plausible that most  hadrons form well outside the interaction region. 
This surprising fact was first realized in the seventies, both 
theoretically and experimentally, when the fallacy of equating a 
nucleus to a ``very dense bubble chamber'' was clearly exposed. We note 
in passing that such is the physical behaviour predicted by the 
Fire-String (FS) model \cite{Angelini}, where the quarks of the 
colliding Nuclei give rise to high mass string-like, unstable 
Hadronic states, the Fire-Strings which decay rather slowly into the 
jet-like structures universally observed in high energy hadronic final 
states. A detailed analysis, to be reported elsewhere \cite{Prep}, 
shows that even in Pb-Pb scattering at very high energies in the 
interaction region there never forms the hot dense matter which 
could (but how?) ``thermalize'' in a QGP, for the high energy density 
of the chromodynamic field that at the time of the collision exists 
in the interaction volume finds its way to the final state only
much later, through a rather slow 
quark-pair creation process. And 
this appears to perfectly agree with a huge body of experimental 
information on the structure of hadronic states \footnote{This can 
also explain the experimentally observed absence of measurable 
correlations between the hadronic jets produced in the decay of 
$W^{+}W^{-}$-pairs at the LEP.}.
\par It is for the above reasons that we have searched for a simpler, more 
realistic explanation of these interesting findings in a completely 
different direction, i.e. in the electromagnetic collective properties of a 
heavy nucleus, that are well known since a long time in the 
thouroughly studied phenomena related to the ``giant resonances'', with 
frequency (A is the mass number) \cite{Bertrand,Greiner1}

\begin{equation}
\omega_{A}\simeq 78 A^{-\frac{1}{3}¥}\quad \hbox{\rm MeV};
\label{eq:qrt omega} 
\end{equation}
and in this paper we wish to present a detailed analysis of the 
$(e^{+}e^{-})$ pair creation process induced by the `` weakly coherent" 
electromagnetic fields that are generated by such ``giant 
resonances''.
\par The paper is organized as follows: Sect.$\mathrm{2}¥$ contains the analysis of the structure
of the coherent e.m. field generated by the coherent charge 
oscillations of the ``giant resonances'', while in Sect.$\mathrm{3}$ we compute 
the $(e^{+}e^{-})$-pair production process in the ``coherent'' e.m. 
fields associated with the ``giant resonances'' of the colliding nuclei. 
The conclusions and outlook comprise Sect.$\mathrm{4}$.

\section{The weak coherent e.m. field associated with the giant 
resonances.}
Let us consider the e.m. vector potential $\vec{A}$ generated by the 
zero-point modes, whose wave number is $|\vec{k}|=\omega_{A}$, the 
frequency of the giant resonance of the nucleus of mass number A, 
approximately given by (\ref{eq:qrt omega}).According to the discussion 
in Ref.\cite{Prep2}, we may write in the Nucleus rest frame the 
vector potential within a coherence domain (CD) of radius 
$R_{CD}=\frac{\pi}{\omega_{A}}$:
\begin{equation}
	 \vec{A}(\vec{x},t)
	 = e^{-i\omega_{A} t}\frac{1} {\left({2\omega_{A}¥ V}
	 \right)^{\frac{1}{2}}}
	 \int 
	 d\Omega_{\vec{k}} e^{i\vec{k}\cdot \vec{ x}}\sum_{r} 
	 \alpha^{(r)}_{\vec{k}}\vec{\epsilon}^{\,(r)}_{\vec{k}}¥
	+c.c.
 	\label{eq:qrt A}
 \end{equation}¥
where $\vec{\epsilon}^{\,(r)}_{\vec{k}}$, (r=1,2) are the 
transverse $(\vec{k}\cdot \vec{\epsilon}^{\,(r)}=0)$ polarization vectors and 
$\alpha^{(r)}_{\vec{k}}$ the associated amplitudes. Our basic idea is 
that the zero point amplitudes, as a consequence of their coupling to 
the coherent oscillations of the nucleons with frequency 
$\omega_{A}¥$, become partly coherent, i.e. can be written as the sum:

\begin{equation}
 	\alpha^{(r)}_{\vec{k}}=\alpha^{(r)}_{\vec{k},c}+
	 \alpha^{(r)}_{\vec{k},inc}
 	\label{eq:qrt Alpha}
 \end{equation}¥
of a coherent and an incoherent part, whose time averages are:

\begin{equation}
<|\alpha^{(r)}_{\vec{k}}|^{2}¥>=|\alpha^{(r)}_{\vec{k},c}|^{2}¥ +
  <|\alpha^{(r)}_{\vec{k},inc}|^{2}¥>=\frac{1}{2¥}.
 	\label{eq:qrt Alpha modulo}
 \end{equation}

We look for a vector potential $\vec{A}(\vec{x},t)$ of the 
particularly simple form:

\begin{equation}
	 \vec{A}(\vec{x},t)
		 = e^{-i\omega_{A} t}a(r)\hat{e}(\theta,\phi)+c.c.
 	\label{eq:qrt A A}
 \end{equation}¥
 where r, $\theta$, $\phi$ are the polar coordinates of the CD 
 centered on the rest frame of the Nucleus, and $\hat{e}(\theta,\phi)$ 
 is a unit vector to be determined by the transversality condition:

 \begin{equation}
 	\vec{\nabla}\cdot\vec{A}=0
 	\label{eq:qrt  gauge}
 \end{equation}¥
 A simple analysis shows that:
 
  \begin{equation}
\hat{e}(\theta, \phi)=(-\sin{\phi},\cos{\phi},0)\label{eq:qrt hat{e}}		
\end{equation}
From the representation (\ref{eq:qrt A}) it requires little analysis to 
show that the coherent part of the amplitudes $\alpha^{(r)}_{\vec{k}}$ 
must have the form  

\begin{equation}
	\alpha^{(r)}_{\vec{k},c}=\lambda \vec{\epsilon}^{\,(r)}_{\vec{k}}
	\hat{e}(\theta, \phi)
	\label{eq:qrt hat{e} alpha¥}
\end{equation}¥
where, according to (\ref{eq:qrt Alpha modulo}), 
$|\lambda|\le\frac{1}{\sqrt{2}}$. The general analysis of 
electrodynamical coherence of Ref.\cite{Prep2} shows that the system 
oscillating Nucleus plus e.m. field stabilizes when the coherent 
amplitudes attain their maximum allowed values, thus we must have
 $|\lambda|=\frac{1}{\sqrt{2}}$. Simple algebra is then required to 
 write for the coherent e.m. field of a single Nucleus in its rest frame:
 
 \begin{equation}
 \vec{A}(\vec{x},t)=\frac{e^{-i\omega_{A}t}}{(2\omega_{A}V_{CD})^{\frac{1}{2}}}	
  \frac{8\pi}{3\sqrt{2}}\frac{\sin\omega_{A}r}{\omega_{A}r}\hat{e}+c.c. ,
  \label{eq:pot vett}
 \end{equation}¥
where

\begin{eqnarray}
 V_{CD}&=&\frac{4\pi}{3}{R}_{CD}^{3} ,\\
 	\nonumber\\
 		R_{CD}&=&\frac{\pi}{\omega_{A}}.    
 \end{eqnarray}¥
 In Table 1
 we report the values of $\omega_{A}$ and $R_{CD}¥$ 
 for the interesting cases of S, Au and Pb.
\par As we have already mentioned, the physical origin and meaning of 
Eq.(\ref{eq:pot vett}) is that the $8\pi$-independent modes of the e.m. 
field of frequency $\omega_{A}$ in the CD, through the interaction 
with the e.m. current generated by the giant resonance's 
fluctuations, get ``aligned'' in a coherent superposition such as 
Eq.(\ref{eq:pot vett}) and in so doing they minimize the total energy of 
the coupled  matter-e.m. field system. As a result we can picture the 
nucleus A as surrounded by a ``photon cloud'' oscillating at the 
frequency of the giant resonance $\omega_{A}¥$ with a well defined 
amplitude (see Eq.(\ref{eq:qrt hat{e} alpha¥})). It should be clear that the 
interaction of the ``photon clouds'' of the Heavy Ions, when they are 
brought to overlap at high energy, in the strong 
supercritical electric field that gets created in the collision, can 
induce the 
production of $e^{+}e^{-}¥$ pairs of a well defined character, which 
we are now going to study.

\begin{figure}
	\centering
     \mbox{\epsffile[0 150 400 650]{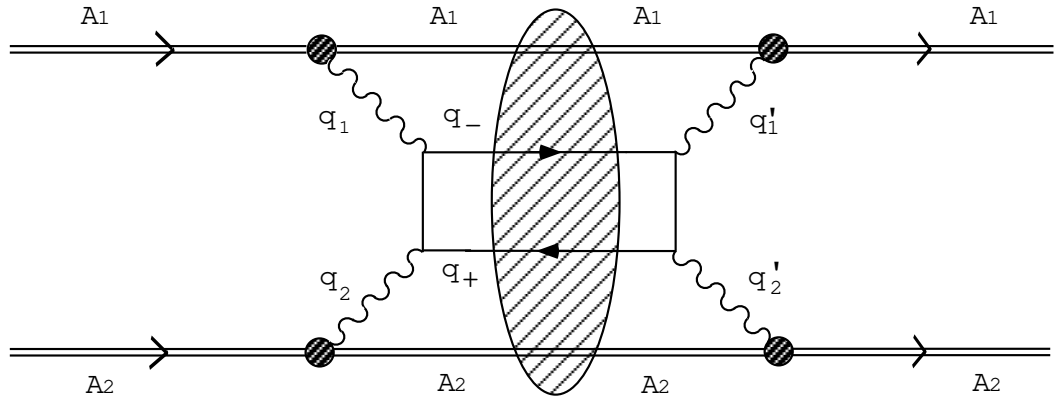}}
	\label{processo tot}
     \caption{\em Feynman graph for the complete process in the Heavy 
     Ion collision.}
\end{figure}

\begin{center}
	{\normalsize { \textbf{III. THE $e^{+}e^{-}$ PAIR PRODUCTION PROCESS.}}}
\end{center}
The process we wish to investigate is depicted in Fig.\ref{processo tot}
where the Heavy Ions $A_{1}¥$ and $A_{2}¥$ surrounded by their ``photon 
clouds'' described by $A_{\mu}(q_{1}¥)¥$ and $A_{\mu}(q_{2}¥)$, the 
Fourier transforms of the boosted e.m. fields of Eq.(\ref{eq:pot 
vett}), collide inelastically, while their ``photon clouds'' produce an 
$e^{+}e^{-}$ pair through the well-known QED mechanism of 
``photon-photon pair creation''.
\par
\begin{figure}
\centering
\mbox{\epsffile[0 0 400 400]{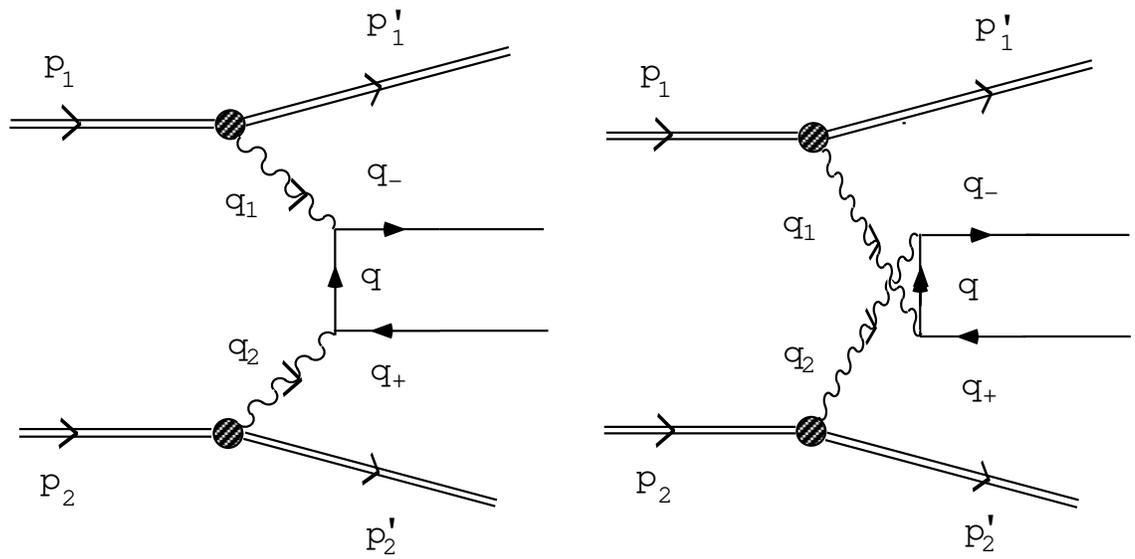}}
\label{fig::Grafici di Feynman}
\vskip 2 cm
\caption{\em Feynman diagram of the $e^{+}e^{-}$ pair QED 
photoproduction mechanism.}
\end{figure}

\begin{figure}\centering	
	 \centering	
	 \mbox{\epsffile[0 0 400 450]{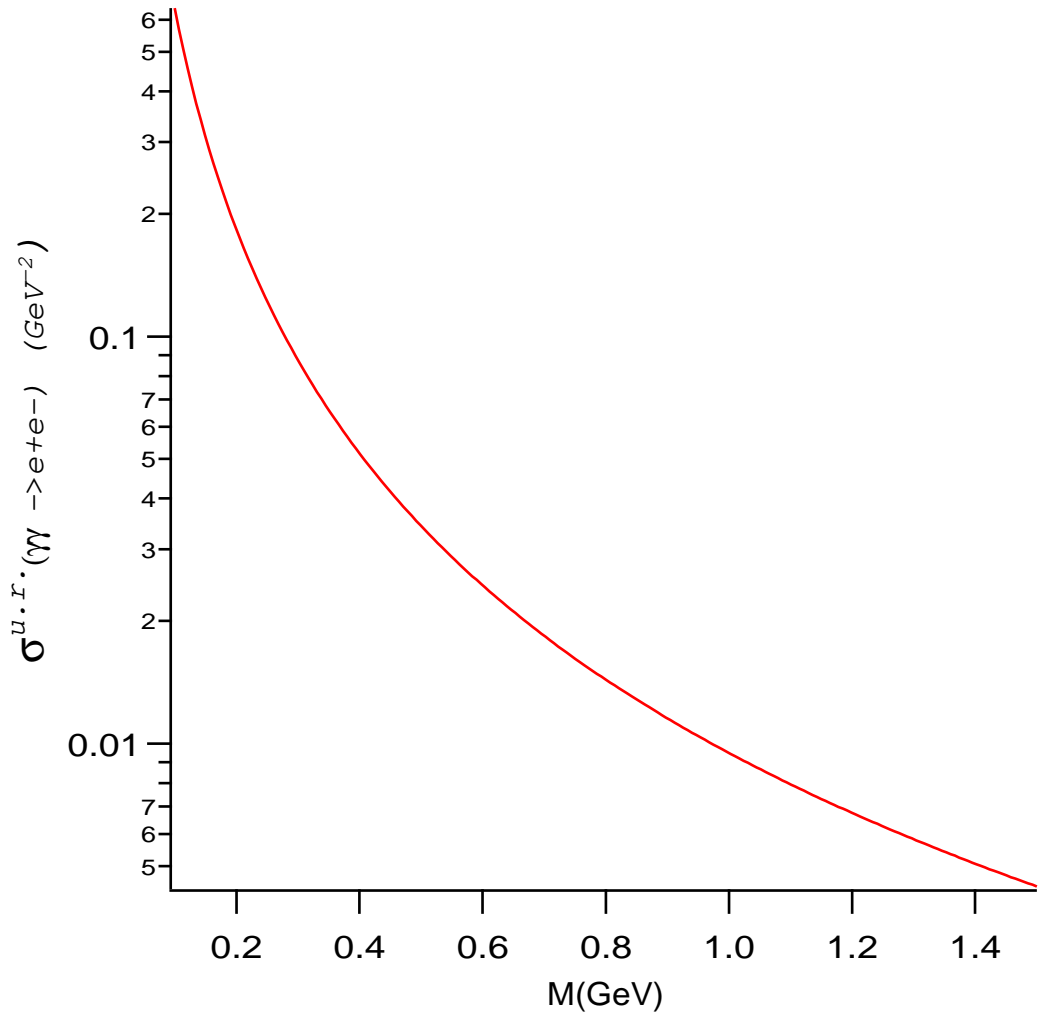}}
	 \vskip 2 cm
	\caption{\em Total unpolarized cross section of the pair production 
	$\gamma\gamma \rightarrow e^{+}e^{-}$ in the energy range of  
	interest for this work.}
	\label{u.r.SIGFREEvsM}
\end{figure}

  From the factorizable structure of the 
diagram in Fig.\ref{processo tot} we can write:

\begin{equation}
	dN_{(e^{+}e^{-}+hadrons)}=dN_{e^{+}e^{-}}*dN_{hadrons}
	\label{eq:art dN}
\end{equation}¥
where the  $e^{+}e^{-}¥$ yield $dN_{(e^{+}e^{-})}$ is calculated by 
squaring the amplitudes associated to the diagrams of
 Fig.\ref{fig::Grafici di Feynman}, which can be written as 
 
 \begin{equation}
S_{rs}¥=\int\frac{d^{4}q_{1}}{(2\pi)^{4}}\frac{d^{4}q_{2}}{(2\pi)^{4}}(2\pi)^{4}
\delta^{4}(q_{1}+q_{2}-q_{-}-q_{+})A_{1\mu}(q_{1})A_{2\nu}(q_{2})T^{\mu\nu}_{rs}
\label{eq::produzione esaltata S}	
\end{equation}
where $T^{\mu\nu}_{rs}$ represents the well known leptonic part of the 
Feynman amplitude. The square of (\ref{eq::produzione esaltata S})
can then be written

\begin{eqnarray}
|S|^{2}&=&\int\frac{d^{4}q_{1}}{(2\pi)^{4}}\frac{d^{4}q_{2}}{(2\pi)^{4}}
\frac{d^{4}q_{1}^{\prime}}{(2\pi)^{4}}\frac{d^{4}q_{1}^{\prime}}{(2\pi)^{4}}
(2\pi)^{4}\delta^{4}(q_{1}+q_{2}-q_{-}-q_{+}) *  \nonumber\\
\nonumber\\
&&* (2\pi)^{4}\delta^{4}(q^{\prime}_{1}+q^{\prime}_{2}-q_{-}-q_{+})A_{1\mu}
(q_{1})A_{2\nu}(q_{2})
A_{1\mu^{\prime}}(q_{1}^{\prime})A_{2\nu^{\prime}}(q_{2}^{\prime})
T_{rs}¥^{\mu\nu}T_{rs}^{{\mu^{\prime}\nu^{\prime}}*}
 \nonumber\\
 \label{eq::produzione 
esaltata |S| QUADRO}
\end{eqnarray}
where we can write for the generic Fourier transformed e.m. field:

\begin{equation}
	A_{\mu}(q)=a(q)\epsilon_{\mu}(q).
	\label{sdfkkljfkjlsfd¥}
\end{equation}¥
From (\ref{eq:pot vett}), in the Nucleus rest frame we have:

\begin{eqnarray}
	a(q_{0},\vec{q})&=
	&\int_{-\frac{\pi}{¥\omega}¥}^{\frac{\pi}{¥\omega}¥}dt 
     e^{iq_{0}t¥}\int_{R_{CD}}¥d^{3}xe^{i\vec{q}\cdot \vec{x}} 
\frac{8\pi}{¥3\sqrt{2}}\frac{e^{-i\omega t}¥}{¥\sqrt{(2\omega 
V_{CD})}}\frac{\sin{\omega r}}{¥\omega r}¥¥¥¥¥
	 \\&=&
	\frac{32\pi^{2}¥}{¥3}\frac{1}{\sqrt{(\omega V_{CD})}}
	\frac{\sin{\frac{\pi}{\omega}}(q_{0}-\omega¥)}
	{(q_{0}¥-\omega)}
	\frac{\sin{(\frac{\pi|\vec{q}\,|}{\omega}})}{|\vec{q}\,|(\vec{q}^{\,2}-\omega^{2})}
	\label{deltapprox}
\end{eqnarray}
Please note that in Fourier transforming we have extended the 
space-time integration to the finite (but large on a QCD scale) domain 
of the CD ($R_{CD}=\frac{\pi}{\omega}¥$). Thus the amplitudes $a(q)$
are highly peaked functions around the values $q_{0}\simeq\omega$ 
and $|\vec{q}\,|\simeq\omega$, and this means that in the 
momentum-integrations in (\ref{eq::produzione 
esaltata |S| QUADRO}) the important regions will be when 
$q_{1}\simeq q_{1}'¥¥$ and $q_{2}\simeq q_{2}'¥¥$, their 
sharpness being controlled by the  extent of ${V_{CD}}$ and 
$T_{CD}=2R_{CD}$ (note that the relative motion of the ions is 
ultrarelativistic). Thus, to lowest order in $\frac{\omega}{q_{N}¥}$ 
(where $q_{N}\simeq O(m_{\pi}¥)¥$, a typical QCD scale), we may derive 
the following approximations:

\begin{eqnarray}
&(2\pi)^{4}&
\delta^{4}(q_{1}+q_{2}-q_{1}'-q_{2}')a(q_{1}¥)a^{*¥}¥(q_{1}'¥)a(q_{2}¥)
a^{*¥}(q_{2}¥')\nonumber\\
&\simeq&
(2\pi)^{4}\delta^{4}(q_{1}+q_{2}-q_{1}'-q_{2}')a(q_{1}¥)a^{*¥}¥(q_{1}'¥)a(q_{2}¥)
a^{*¥}(q_{2}¥')\nonumber\\&&(2\pi)^{4}\frac{\delta^{4}(q_{1}-q_{1}')}
{V_{1}T_{1}¥¥¥}¥
(2\pi)^{4}\frac{\delta^{4}(q_{2}-q_{2}')}{V_{2}T_{2}¥¥¥}\\
&\simeq& (2\pi)^{4}\frac{\delta^{4}¥(q_{1}-q_{1}')}{V_{1}T_{1}¥}(2\pi)^{4}
\frac{\delta^{4}¥(q_{2}-q_{2}')}{V_{2}T_{2}¥¥¥}|a(q_{1}¥)|^{2}|
a(q_{2}¥)|^{2}VT\nonumber\\
&\simeq& 
(2\pi)^{4}\frac{\delta^{4}¥(q_{1}-q_{1}')}{\sqrt{(V_{1}T_{1})}}
\frac{\delta^{4}¥(q_{2}-q_{2}')}{\sqrt{(V_{2}T_{2})}}
|a(q_{1}¥)|^{2}|a(q_{2}¥)|^{2}\nonumber
\end{eqnarray}¥
where 

\begin{equation}
	V_{i}T_{i}=\frac{4\pi}{3}R^{3}_{CD_{i}¥}2R_{CD_{i}¥}=\frac{8\pi}{¥3}
	\left(\frac{\pi}{¥\omega_{i}¥}¥\right)^{4}¥
	\label{kdkkdkdk}
\end{equation}¥
$i=1,2$, and we have approximated the interaction space-time domain:

\begin{equation}
	VT \simeq [V_{1}T_{1}V_{2}T_{2}]^{\frac{1}{2¥}}
	\label{™pwpof}
\end{equation}¥
It is now straightforward to derive for the differential 
$e^{+}e^{-}¥$ yield the following expression:

\begin{equation}
dN_{e^{+}e^{-}}=\int\frac{d^{4}q_{1}}{(2\pi)^{4}}\frac{d^{4}q_{2}}{(2\pi)^{4}}
\frac{6\omega^{2}_{1}\omega^{2}_{2}}{\pi(2\pi)^{4}}
|a(q_{1})|^{2}|a(q_{2})|^{2}2M^{2}d\sigma_{\gamma\gamma \rightarrow 
e+e-}\label{eq::produzine esaltata sigma dNee}
\end{equation}
where $a(q_{1}¥)$ and  $a(q_{2}¥)$ are the 4-dimensional Fourier 
transforms of the boosted ``photon clouds'', and $d\sigma_{\gamma\gamma \rightarrow 
e^{+}e^{-}}$ is the well-known differential cross section of 
$\gamma\gamma$ annihilation in $e^{+}e^{-}¥$ pairs of mass M. Fig.\ref{u.r.SIGFREEvsM} 
 reports the shape of $\sigma^{u.r.}_{\gamma\gamma \rightarrow e^{+}e^{-}}(M)$ as a 
 function of M \cite{Greiner2}, where $\sigma^{u.r.}_{\gamma\gamma \rightarrow e^{+}e^{-}}(M)$
 is the total cross section $\sigma_{\gamma\gamma \rightarrow e^{+}e^{-}}(M)$
 in the ultrarelativistic limit. A simple approximate computation of 
 Eq.(\ref{eq::produzine esaltata sigma dNee}) employs again the 
 $\delta$-like approximation:
 
 \begin{equation}
	a(q)^{2}=\mu \delta(\frac{P\cdot q}{m}-\omega)
	\delta(q^{2})
	\label{pssd}
\end{equation}¥
where $\mu$ is the approximate normalization factor, and P is the 
Nucleus 4-momentum. Thus in the rest frame $(P=(m,\vec{0}))$ 

\begin{equation}
	a(q)^{2}=\mu \delta(q_{0}-\omega) \delta(\omega^{2}-\vec{q}^{2}¥),
		\label{ldf¥}
\end{equation}¥
approximating the $q_{0}¥$- and $\vec{q}$- Fourier transforms of 
(\ref{eq:pot vett}), both of which are steeply peaked at $\omega$, the 
giant resonance frequency. With these simple approximations from 
(\ref{eq::produzine esaltata sigma dNee}) we derive:

\begin{equation}
	\frac{d^{2}N_{e^{+}e^{-}}}{dMdy}=\frac{32}{3\pi^{5}}
	\frac{1}{\omega_{1}\omega_{2}}
	{\mathcal F} (M^{2},y)M^{3}\sigma_{\gamma\gamma \rightarrow 
	e^{+}e^{-}}(M),
	\label{qeqeqeqre}
\end{equation}¥
where the `` form-factor" ${\mathcal F} (M^{2},y)$, ($y$ is the pair rapidity 
in the CM), is given by

\begin{eqnarray}
{\mathcal F}(M^{2},y)&=&\int d^{4}q_{1}d^{4}q_{2}\quad \delta(q_{1}^{2})\delta(q_{2}^{2})
\delta(\frac{P_{1}\cdot q_{1}}{m_{1}}-\omega_{1})
\delta(\frac{P_{2}\cdot q_{2}}{m_{2}}-\omega_{2})*\nonumber\\
\nonumber\\
&&*\delta[(q_{1}+q_{2})^{2}-M^{2}]\delta(y-\frac{1}{2}\ln R)
\label{eq_teta teta}
\end{eqnarray}
where

\begin{equation}
R=\frac{q_{01}+q_{02}+q_{z1}+q_{z2}}{q_{01}+q_{02}-q_{z1}-q_{z2}}
\label{eq_teta RR}
\end{equation}
The calculation of the `` form factor" ${\mathcal F}(M^{2},y)$ can be 
best carried out by means of the Sudakov decomposition:

\begin{eqnarray}
q &=& \alpha n+\beta \overline{n}+q_{\bot}¥, \label{eq kdk}\\
n &\equiv &\frac{\sqrt{s}}{2}(1,0,0,1), \label{eq n} \\
\overline{n} &\equiv & \frac{\sqrt{s}}{2}(1,0,0,-1) \label{eq n bar}
\end{eqnarray}
$\sqrt{s}$ being the CM-energy of the colliding nucleons in the 
ultrarelativistic limit, and:

\begin{equation}
\int{d^{4}q}=\frac{s}{2}\int{d\alpha d\beta d^{2}q_{\bot}}
\end{equation}
The result, which can be obtained in a straightforward manner, is 
$(\omega_{1}\ge\omega_{2}¥)$

\begin{eqnarray}
{\mathcal F}(M^{2},y)&=&\frac{\pi^{2}m^{2}¥¥}{¥s}¥\theta 
\left[4\omega_{1}¥\omega_{2}¥-\frac{M^{2}m^{2}¥¥}{¥s}\right]¥*\nonumber\\
&*&\theta \left[\ln \frac{2\omega_{1} \sqrt{s}¥}{¥mM}-y \right]
\theta \left[y-\ln \frac{Mm}{2\omega_{2}\sqrt{s}¥¥}¥\right] \label{dfgsdfgr}
\end{eqnarray}
where m is the nucleon mass and $\sqrt{s}$ is the CM energy of the 
nucleon-nucleon collision.
\par It is clear that approximating sharp functions with 
$\delta$-functions has produced the simple, but somewhat unrealistic 
result (\ref{dfgsdfgr}), yielding a sharp cut-off at 
$M^{2}=\frac{4\omega_{1}\omega^{2}}{m^{2}¥}s=\frac{8\omega_{1}\omega^{2}E}{m}$, where
 E is the Laboratory energy of the impinging nucleon, $s\simeq2mE$. A 
 more realistic and better approximation can be obtained by letting:
 
 \begin{equation}
\delta(q_{0}-\omega)=\delta(\frac{P\cdot 
q}{m}-\omega)\rightarrow\left({\frac{\lambda}{\pi}}\right)^{\frac{1}{2}} 
e^{-\lambda {(\frac{P\cdot q}{m}-\omega)}^{2}}\label{eq:5}
\end{equation}
i.e. by replacing the $\delta$-like approximation by a Gaussian whose 
parameter $\lambda$ can be evaluated by equating the first-order 
expansion in $(q_{0}-\omega)^{2}¥$ of (\ref{eq:5}) and of
$[\frac{\sin{\frac{\pi}{\omega}}(q_{0}-\omega¥)}{(q_{0}¥-\omega)}]^{2}¥$, 
 (see Eq.(\ref{deltapprox})). Thus we get 
 
 \begin{equation}
\lambda \simeq\frac{1}{¥6}¥\left(\frac{\pi}{¥\omega}¥
\right)^{2}
	\label{eq lambda sopr}
\end{equation}¥
In this way the `` form-factor" ${\mathcal F}(M^{2},y)$ is given by 
(neglecting a weak rapidity dependence):

 \begin{eqnarray}
{\mathcal F}(M^{2},y)&=&{\frac{(\lambda_{1}\lambda_{2})^{\frac{1}{2}}}{\pi}}\int\int 
d\Omega_{1}¥ d\Omega_{2}\quad 
e^{-\lambda_{1}¥(\omega_{1}¥-\Omega_{1}¥)^{2}} 
e^{-\lambda_{2}¥(\omega_{2}¥-\Omega_{2}¥)^{2}}
*\nonumber\\ \nonumber\\ &&* \left[\frac{\pi^{2}m^{2}}{s¥}¥
\theta \left( \frac{s'}{m^{2}} 
4\Omega_{1}\Omega_{2} -\frac{m^{2}¥M^{2}}{s}¥\right) 
\right].\label{eq:11} 
\end{eqnarray}
An approximate evaluation of this integral yields the following 
simple result:

\begin{equation}
 	{\mathcal F} (M^{2}) =\pi^{\frac{3}{2}}\frac{m^{2}}{s}
	 \int_{-\sqrt{\frac{\lambda_{1}\lambda_{2}}{\lambda_{1}\omega_{1}^{2} 
	 +\lambda_{2}\omega_{2}^{2}}} 
	 \left(\omega_{1}\omega_{2}-\frac{m^{2}M^{2}}{4 s} 
	 \right)}^{+\infty} \quad e^{-t^{2}} dt.
 	\label{s"s"d"d"}
 \end{equation}¥
 \section{Conclusions and outlook.}
\begin{figure}[thp]
 \centering
     \mbox{\epsffile[0 0 400 500]{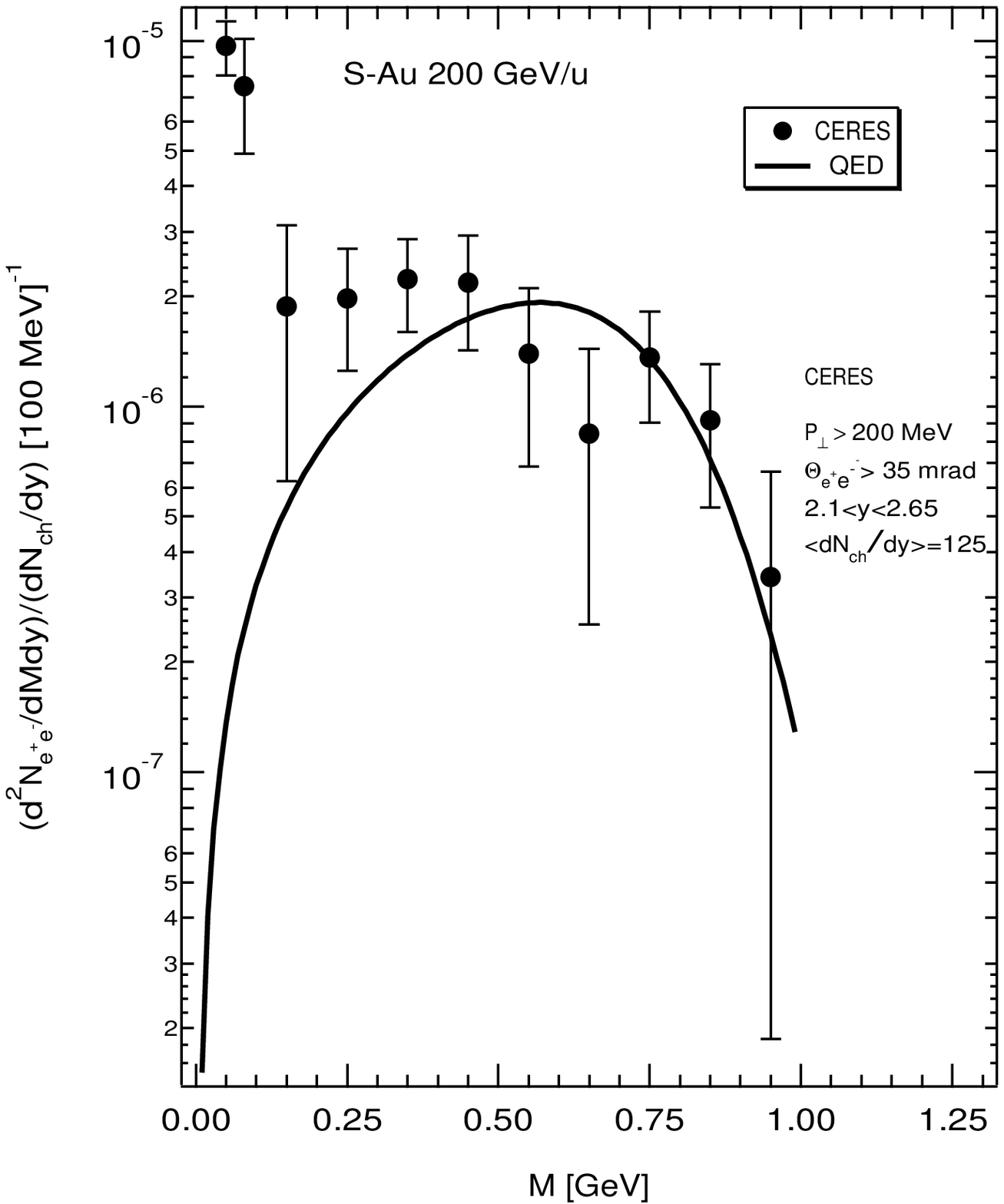}}
     \vskip 1.5 cm
 \caption{\em Normalized differential spectrum of the photoproduced $e^{+}e^{-}$
 pairs calculated in this work and compared with the experimental data 
 for S-Au of Ref.3.}
 \label{fig::conclus1 fig S-Au teor}
 \end{figure}

\begin{figure}[thp]
     \centering
     \mbox{\epsffile[0 0 400 500]{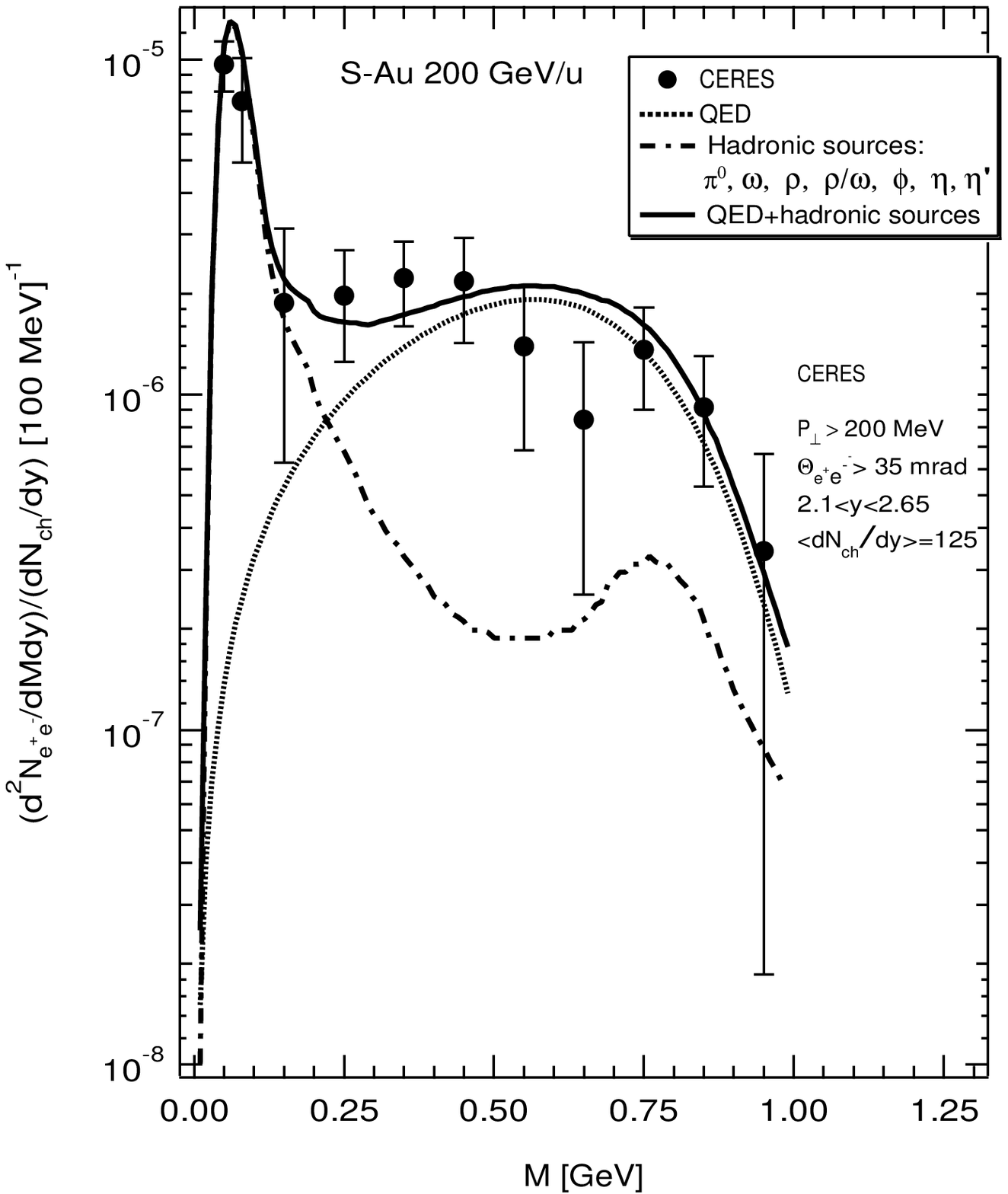}}
     \vskip 1.5 cm
\caption{\em Normalized differential spectra of the $e^{+}e^{-}$ pair 
 production (QED, hadron sources, QED+hadron sources) compared with 
 the experimental data for S-Au of Ref.3.}
 \label{fig::conclus2 fig S-Au teor}
 \end{figure}

\begin{figure}[thp]
\centering
 \mbox{\epsffile[0 0 400 500]{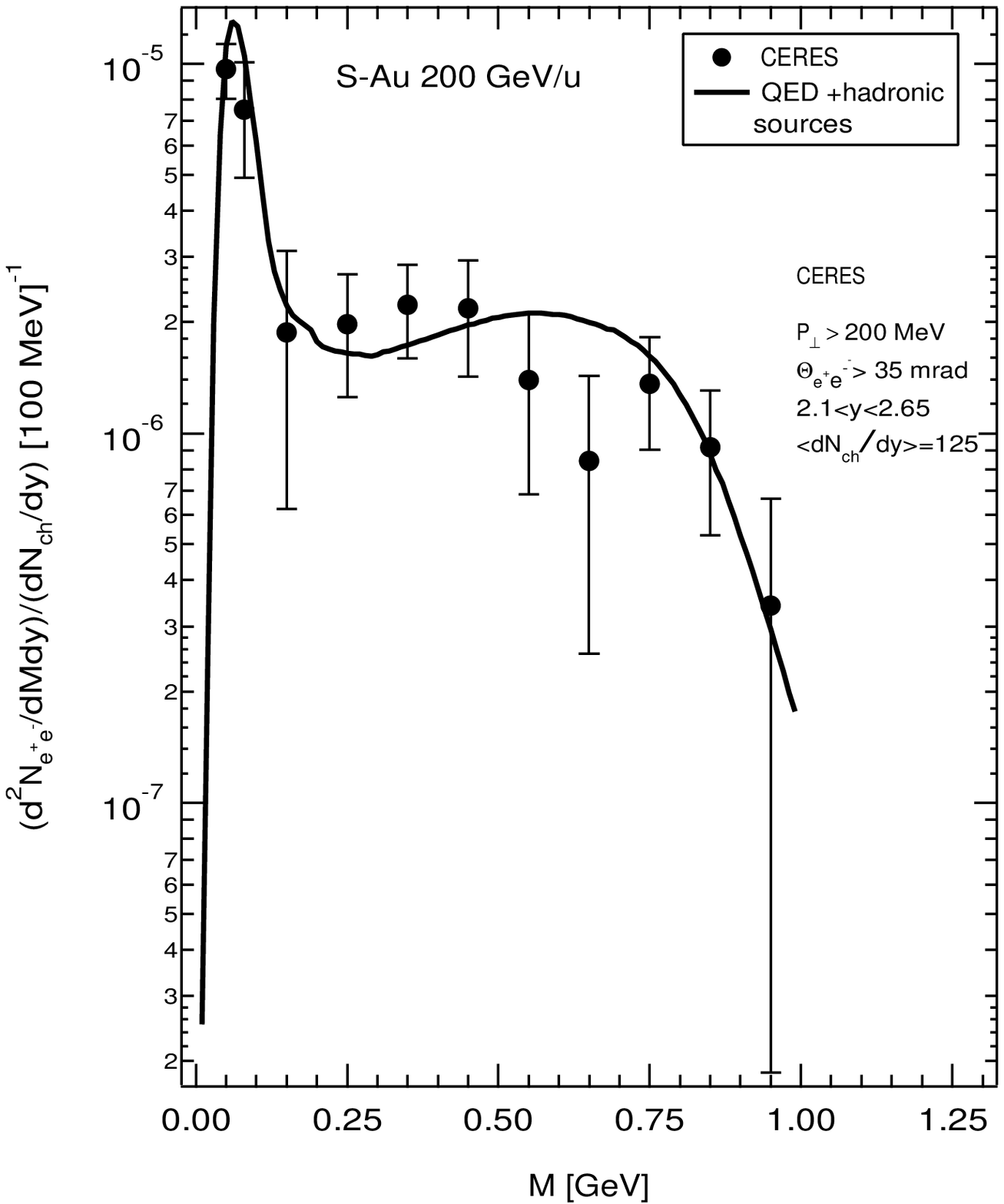}}
 \vskip 1.5 cm
\caption{\em Normalized differential spectrum of the entire $e^{+}e^{-}$ 
pair production (QED+hadron sources) compared with the experimental 
data for S-Au of Ref.3.}
\label{fig::conclus3 fig S-Au teor}
\end{figure}

\begin{figure}[thp]
\centering
 \mbox{\epsffile[0 0 400 500]{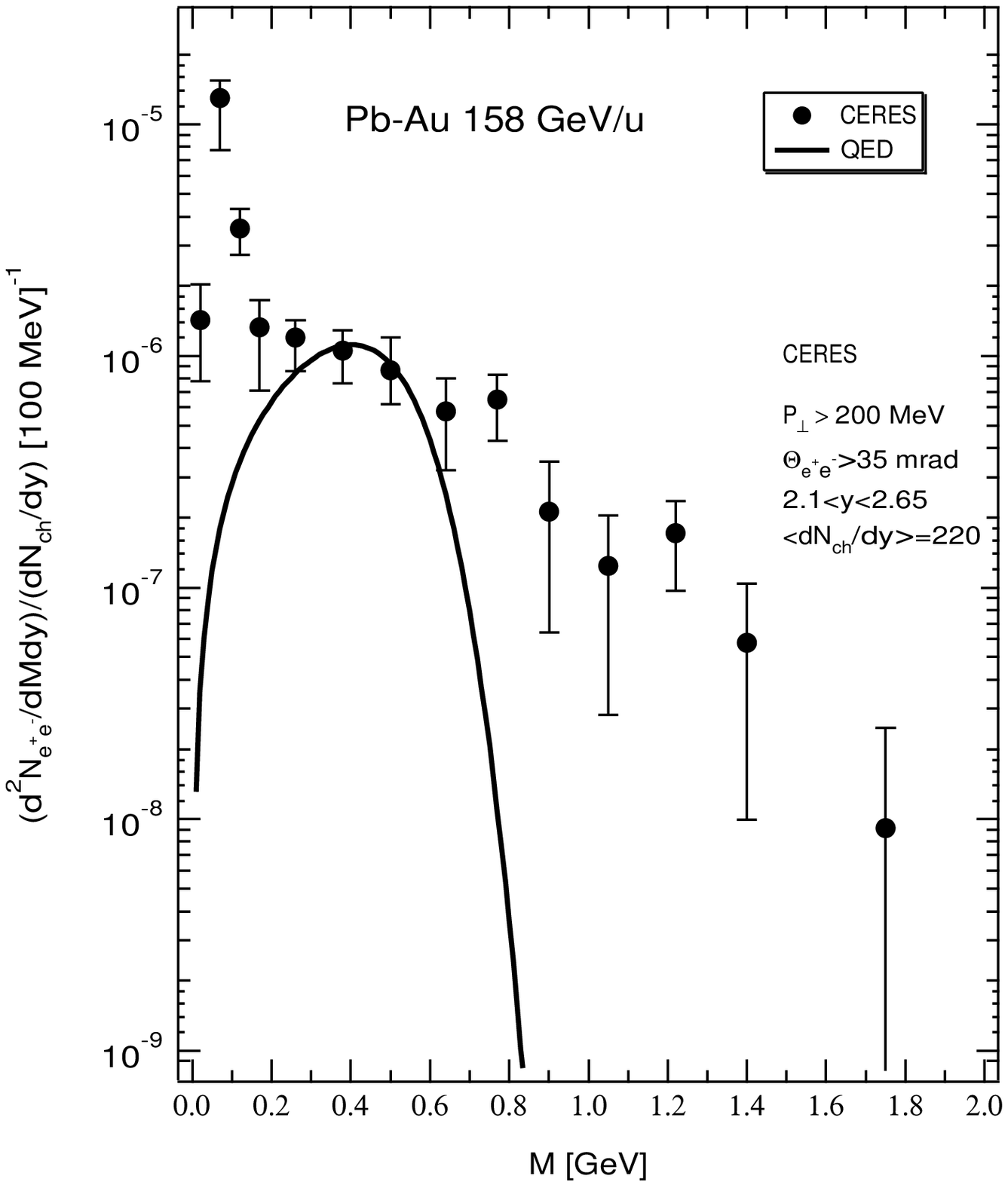}}
 \vskip 1.5 cm
\caption{\em Normalized differential spectrum of the photoproduced $e^{+}e^{-}$
pairs calculated in this work and compared with the experimental data 
for Pb-Au of Ref.5.}
\label{fig::conclus1 fig Pb-Au teor}
\end{figure}

\begin{figure}[thp]
\centering
 \mbox{\epsffile[0 0 400 500]{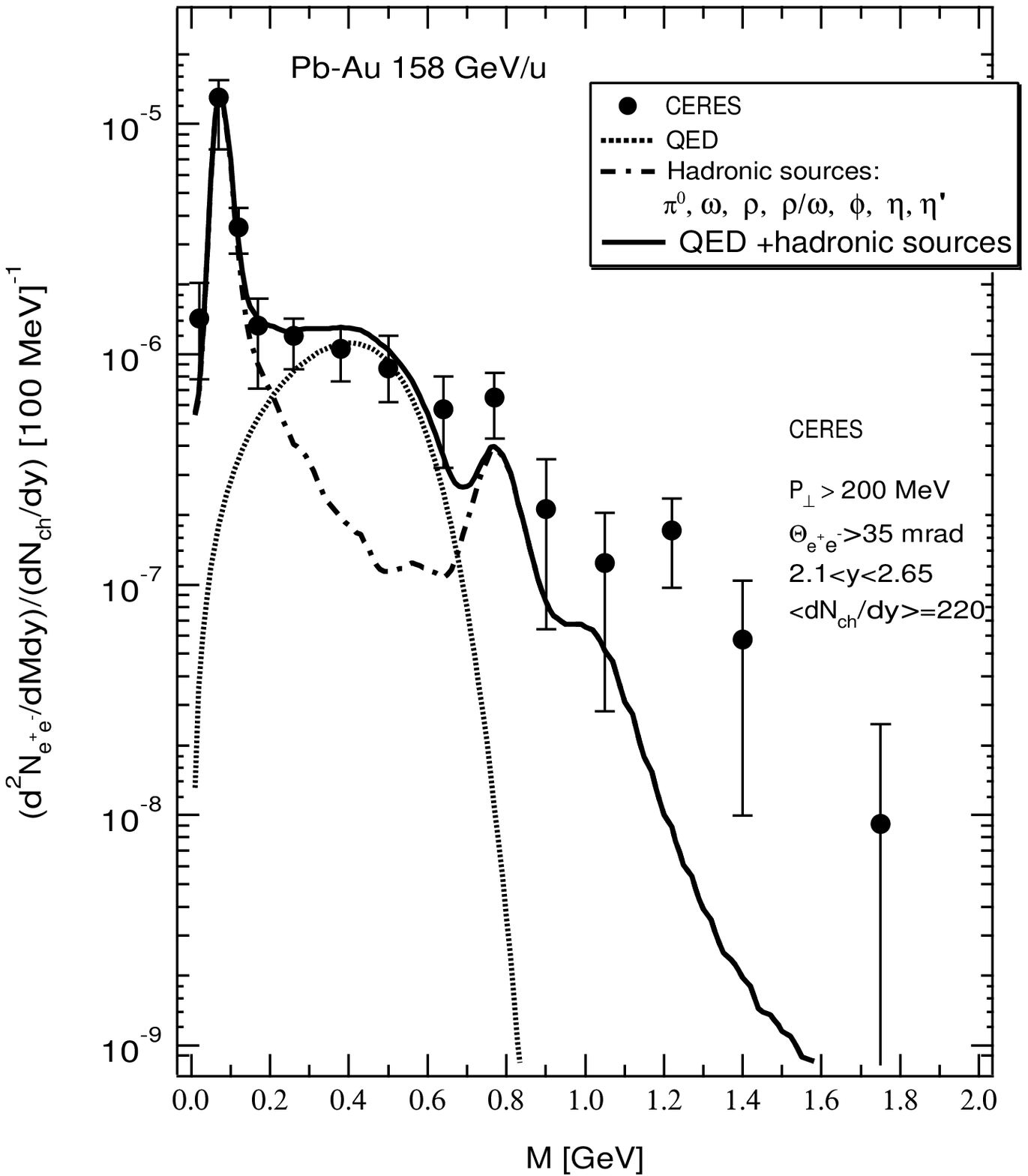}}
 \vskip 1.5 cm
\caption{\em Normalized differential spectra of the $e^{+}e^{-}$ pair 
production (QED, hadron sources, QED+hadron sources) compared with 
the experimental data for Pb-Au of Ref.5.}
\label{fig::conclus2 fig Pb-Au teor}
\end{figure}

\begin{figure}[thp]
\centering
 \mbox{\epsffile[0 0 400 500]{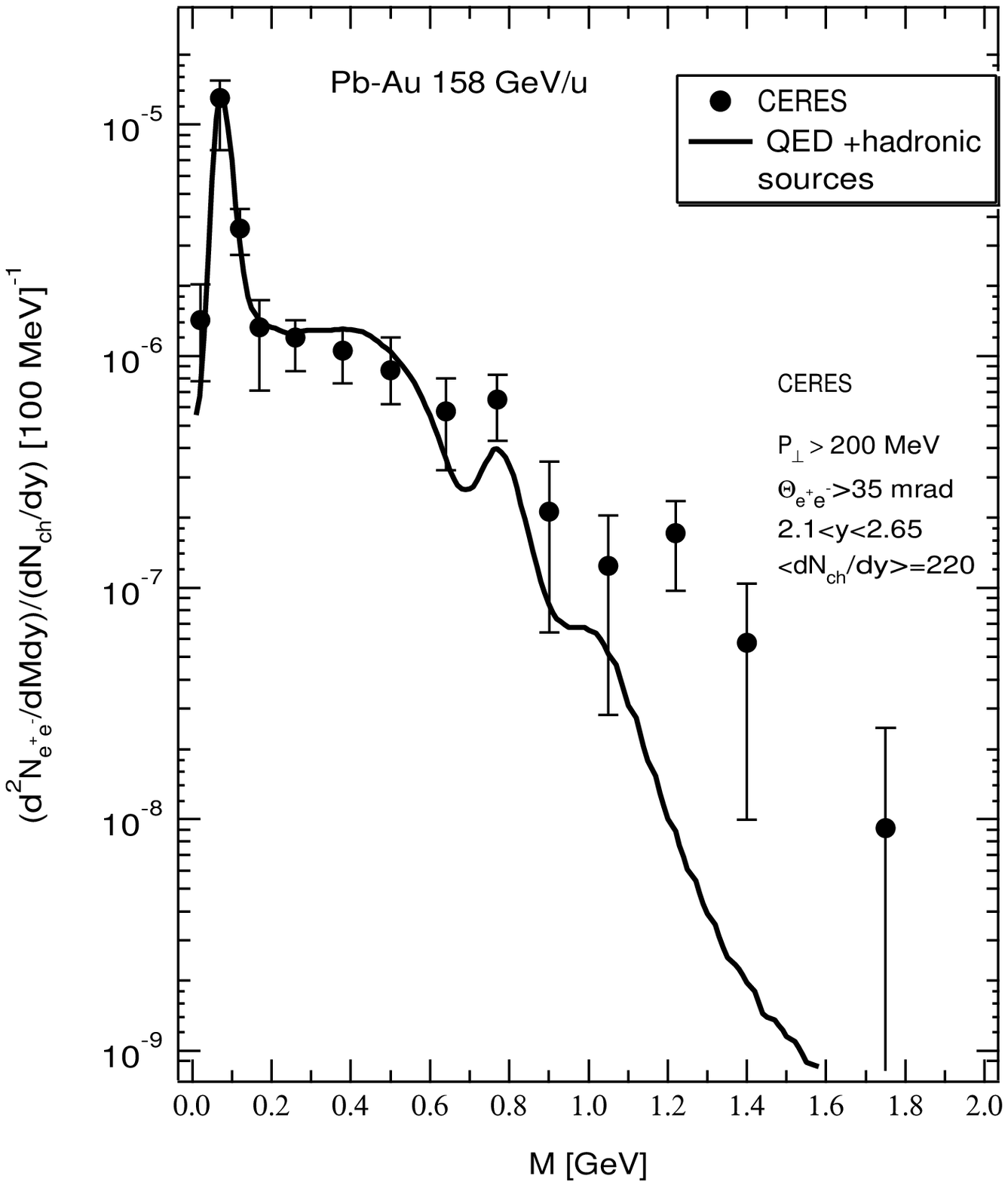}}
 \vskip 1.5 cm
\caption{\em Normalized differential spectrum of the entire $e^{+}e^{-}$ 
pair production (QED+hadron sources) compared with the experimental 
data for Pb-Au of Ref.5.}
\label{fig::conclus3 fig Pb-Au teor}
\end{figure}

\begin{figure}[thp]
\centering
 \mbox{\epsffile[0 0 400 500]{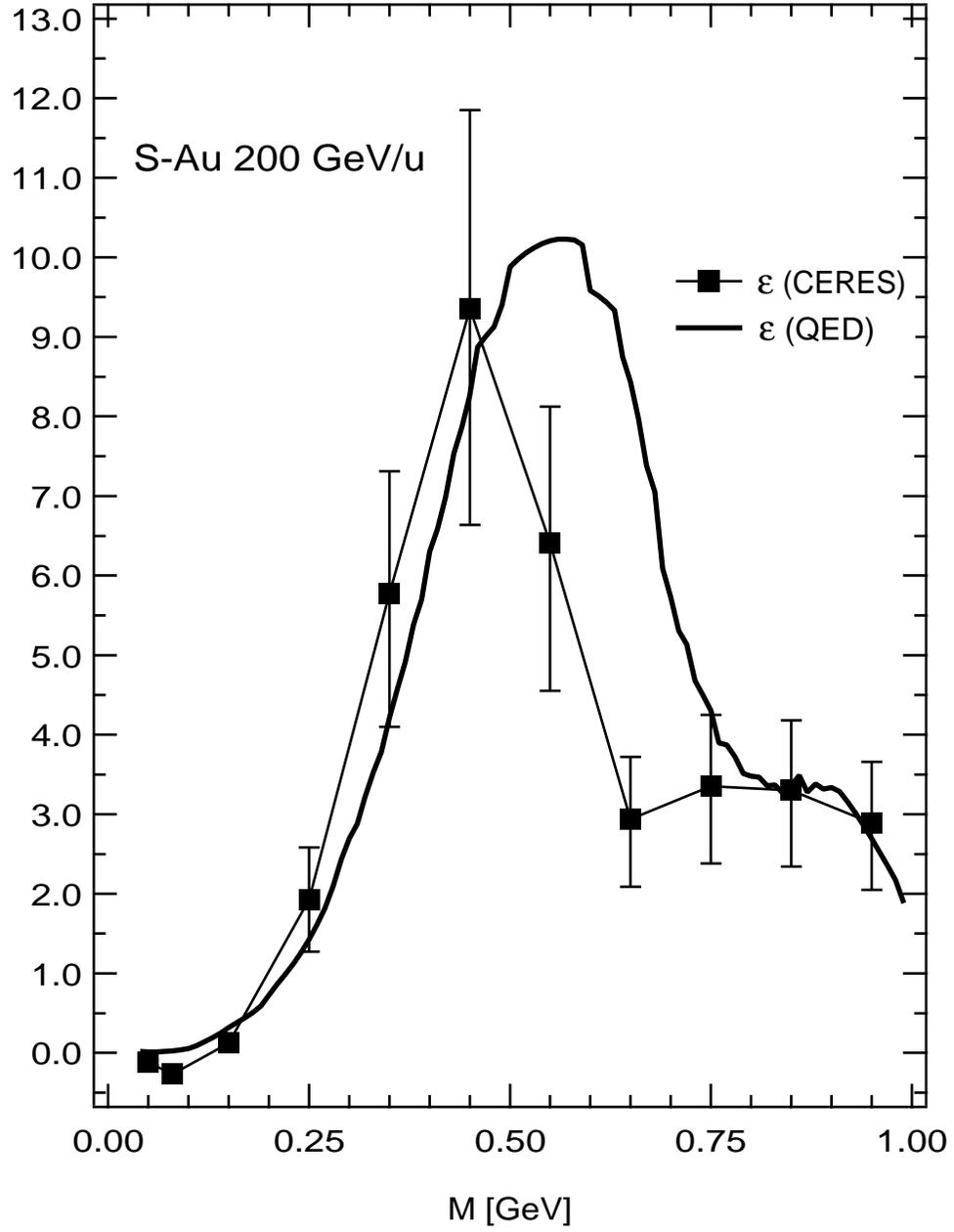}}
 \vskip 1.5 cm
 \caption{\em Comparison of the experimental S-Au (CERES, Ref.3)
 $e^{+}e^{-}$ pairs relative excess $\epsilon_{exp}$ with our 
 theoretical  $\epsilon_{th}$  relative excess calculated in this work.}
\label{fig::conclus fig S-Au teor}
\end{figure}

\begin{figure}[thp]
\centering
 \mbox{\epsffile[0 0 400 500]{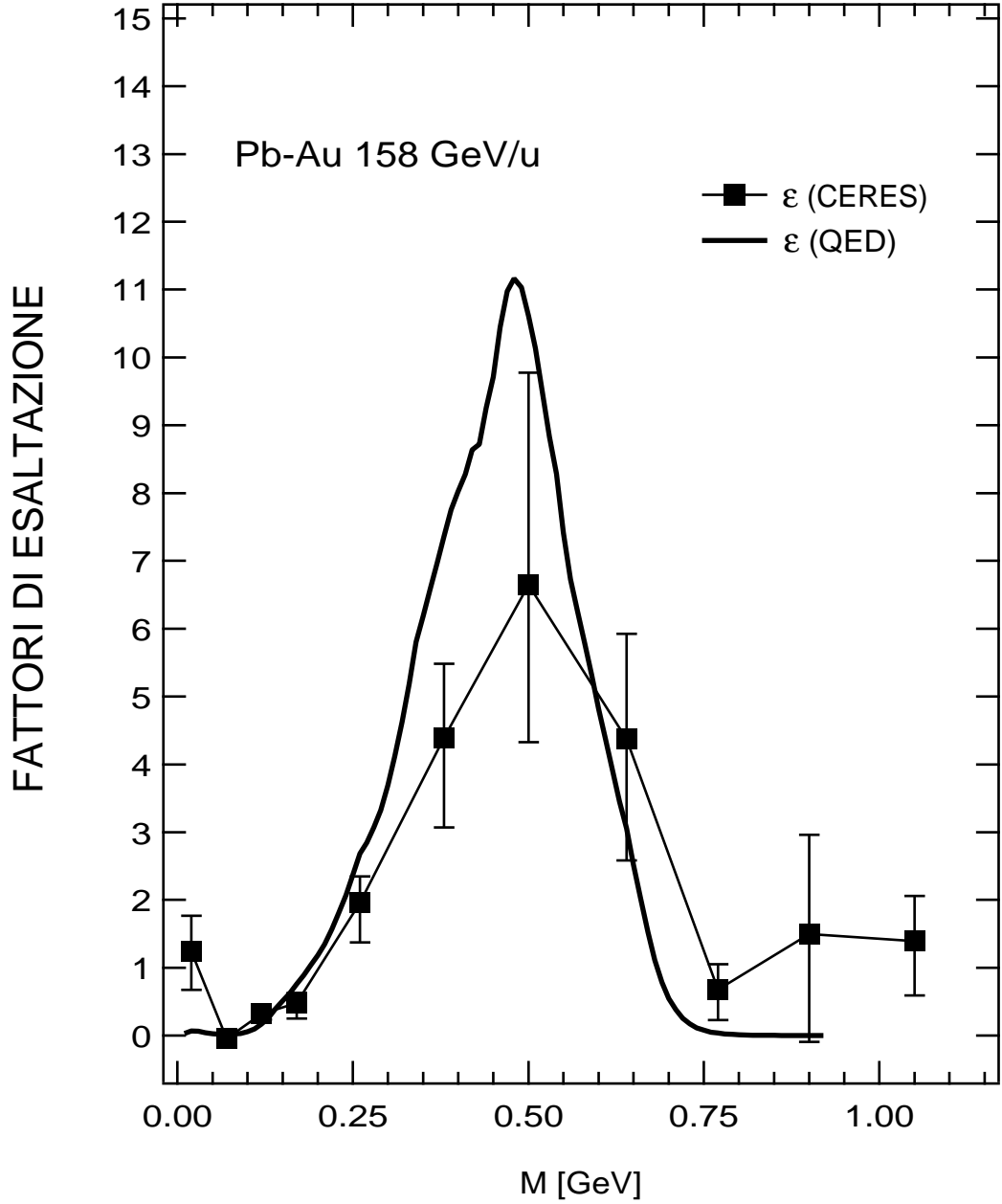}}
 \vskip 1.5 cm
\caption{\em Comparison of the experimental Pb-Au (CERES, Ref.5)
 $e^{+}e^{-}$ pairs enhancement factor $\epsilon_{exp}$ with our 
 $\epsilon_{th}$ enhancement factor calculated in this work.}
\label{fig::conclus fig Pb-Au teor}
\end{figure}

 By feeding in the appropriate values of the frequencies $\omega$ and 
 of the Lab-energies E in the formulae derived in the preceding 
 Section, our results are compared with the experimental information 
 in Figs.\ref{fig::conclus1 fig S-Au teor},\ref{fig::conclus2 fig S-Au teor},\ref{fig::conclus3 fig S-Au teor},\ref{fig::conclus1 fig Pb-Au teor},
\ref{fig::conclus2 fig Pb-Au teor},\ref{fig::conclus3 fig Pb-Au teor}, 
for S-Au scattering at $200$ GeV per 
 nucleon and Pb-Au at $158$ GeV per nucleon respectively.\
 A good way to appreciate the success of our calculation is to compare 
 the quantity:
 
 \begin{equation}
 \epsilon_{exp}=\frac{\frac{d^{2}¥N_{ee}¥}{dMd\eta}|_{exp}-
 \frac{d^{2}¥N_{ee}¥}{dMd\eta}|_{known sources}}
 {\frac{d^{2}¥N_{ee}¥}{dMd\eta}|_{known sources}}¥
 \end{equation}
 with
 
\begin{equation}
 \epsilon_{th}=\frac{\frac{d^{2}¥N_{ee}¥}{dMd\eta}|_{th}}
 {\frac{d^{2}¥N_{ee}¥}{dMd\eta}|_{known sources}}¥
 \end{equation}
 This is done for S-Au in Fig.\ref{fig::conclus fig S-Au teor} and 
 for Pb-Au in Fig.\ref{fig::conclus fig Pb-Au teor}. In view of the parameterless nature of our calculation and of the
approximations involved in it, we regard the agreement between theory and
experiments really remarkable.
\par What physics lessons can we learn from all this? The first is that at very high energies the collisions between heavy ions can
still provide us with interesting informations about the collective nuclear
dynamics, and the $e^{+}e^{-}$ pair production at low masses $(M\leq 1$ GeV)
appears to bear witness to the importance and relevance of the collective
excitations of heavy Nuclei in the generation of strong, coherent e.m. fields
in which low-mass pair creation takes place. The conceptual as well as the
calculational simplicity of our work is, in our view, a strong support of the
validity of our physical explanation for the surprising $e^{+}e^{-}$ excess
observed by the mentioned experiments \cite{Aga1,Aga2}. 
\par  Another equally important and relevant lesson is that another possible
signature for the \\``mythical" QGP can be, most likely, explained away, our
computation leaving essentially no room for the $\pi \pi$-annihilation in low
mass $e^{+}e^{-}$ pairs, that should characterize a hot, dense medium such as
QGP. Based on the discussion developed in the Introduction, this fact 
represents a relevant corroboration of the fundamental ideas of QCD, that relegate all 
phenomena of hadroproduction in the final state to a
long-distance, basically slow dynamics where the strong colour fields, created
by the rearrangements of the initial quarks in Fire-Strings \cite{Angelini}, get
slowly converted into hadronic matter, which therefore appears in the
laboratory quite far (several Fermis at high energy)  from the interaction
region. Naturally a good experimental evidence for the QGP, invalidating ideas
that have sprung from decades of hadronic physics, would pose several
fundamental problems not only to the FS-picture but also to the apparent lack
of the mechanisms needed to ``thermalize" the QGP in the relatively short times
involved in high energy heavy-ion scattering. And, in this light, we are
particularly pleased by the results of this work.
\section*{Acknowledgments}
We are indebted to dr. A.Drees for supplying us with the numerical data 
of the CERES S-Au measurements.

\newpage

\begin{center}
	Table 1
\end{center}
\vskip 3 cm
\begin{center}
	\begin{tabular}{|c|c|c|c|}
		\hline
	Nucleus  & A  & $\omega$ $[MeV]$  & $R_{CD}$ $[fm]$   \\
		\hline		\hline
	Au	     & 32    & 23   & 27 \\
		\hline
	S        & 197   & 13.4 &  46 \\
		\hline
	Pb       & 207   & 13.1 & 47   \\
		\hline
	\end{tabular}
\end{center}



\listoffigures

\end{document}